\documentclass[12pt]{article}

\pdfoutput=1

\usepackage{amssymb,graphicx}
\usepackage[intlimits]{amsmath}

\def\td{\tilde }
\def\p{\partial }
\def\w{\omega}

\begin{document}




\renewcommand{\thefootnote}{\fnsymbol{footnote}}
\setcounter{footnote}{0}

\def\caln{\mathcal{N}}
\def\cL{\mathcal{L}}

\begin{titlepage}

\begin{flushright}\footnotesize

\texttt{ICCUB-18-010}
\vspace{0.3cm}
\end{flushright}

\vspace{1.5cm}
\begin{center}

{\large \bf  Exact gravitational plane waves and two-dimensional gravity}

\vspace{0.8cm}

\textrm{Jorge~G.~Russo}\footnote{ jorge.russo@icrea.cat}
\vspace{4mm}

\textit{\footnotesize  Instituci\'o Catalana de Recerca i Estudis Avan\c cats (ICREA), \\
Pg. Lluis Companys, 23, 08010 Barcelona, Spain.}\\
\textit{\footnotesize  Department de Fisica Cuantica i Astrofisica \\
and Institut de Ci\`encies del Cosmos,  \\
Universitat de Barcelona, Mart\'\i \ Franqu\`es, 1, 08028 Barcelona, Spain.}

\vspace{3mm}


\par\vspace{0.4cm}

\textbf{Abstract} \vspace{3mm}

\begin{minipage}{13cm}

We discuss dynamical aspects of gravitational plane waves in Einstein
theory with massless scalar fields. The general analytic solution describes 
colliding gravitational waves with constant polarization, which interact with
scalar waves and, for generic initial data, produce a spacetime singularity at the focusing hypersurface.  There is, in addition,  an infinite family of regular solutions and  an intriguing static geometry supported by scalar fields. Upon dimensional reduction, the theory can be viewed as an exactly solvable two-dimensional gravity model. 
This provides a new viewpoint on the gravitational dynamics.
Finally, we comment on a simple mechanism by which short-distance  corrections in the two-dimensional model can remove the singularity.

\end{minipage}

\end{center}

\vspace{0.5cm}

\end{titlepage}


\newpage
\setcounter{page}{1}
\renewcommand{\thefootnote}{\arabic{footnote}}
\setcounter{footnote}{0}





\def\Xint#1{\mathchoice
   {\XXint\displaystyle\textstyle{#1}}%
   {\XXint\textstyle\scriptstyle{#1}}%
   {\XXint\scriptstyle\scriptscriptstyle{#1}}%
   {\XXint\scriptscriptstyle\scriptscriptstyle{#1}}%
   \!\int}
\def\XXint#1#2#3{{\setbox0=\hbox{$#1{#2#3}{\int}$}
     \vcenter{\hbox{$#2#3$}}\kern-.52\wd0}}
\def\ddashint{\Xint=}
\def\dashint{\Xint-}

\newcommand{\bea}{\begin{eqnarray}} \newcommand{\eea}{\end{eqnarray}}

\newcommand{\be}{\begin{equation}}
\newcommand{\ee}{\end{equation}}

\newcommand\cev[1]{\overleftarrow{#1}}



An important problem in General Relativity (GR) is understanding detailed aspects of its highly non-linear
dynamics and the time evolution of general distributions of matter. This includes, in particular, problems involving the emergence of spacetime singularities, gravitational plane wave collisions and black hole formation.
A complete account of back-reaction in  general
time-dependent configurations is typically very complicated, but there are cases 
where the exact  solution is known.

Since the classic works by Einstein and Rosen \cite{Einstein:1937qu} and  Bondi, Pirani and Robinson \cite{Bondi:1958aj}, many examples of exact gravitational plane waves  have been discovered in the literature and
investigated in great detail (see {\it e.g.} \cite{Khan:1971vh}-\cite{Stephani:2003tm} and references therein). Plane waves play an important role in physics and,  recently, the LIGO experiment led to a major breakthrough by the  direct detection of gravitational waves. Unlike electromagnetic or sound waves, because of the non-linearity of general relativity 
one cannot just add individual gravitational waves to build up the resulting geometry.
This note is aimed to discuss a new perspective on the problem.

Our starting point is Einstein gravity coupled to 
free massless scalar fields,
\be
S = \frac{1}{2\kappa^2} \int d^4 x \sqrt{-G} \left(R^{(4)} -2\p_\mu\phi^I \p^\mu \phi^I \right)\ ,\qquad I=1,...,N\ .
\nonumber
\ee
Note that we are using a non-canonical normalization for the scalar fields.
Next, we consider the ansatz
\be
ds^2=g_{ij}(x,t) dx^i dx^j +e^{-2\Phi (x,t)} (e^{2\varphi } dy^2 +e^{-2\varphi }  dz^2)\ ,\qquad \phi^I =\phi^I(x,t)
\ .
\nonumber
\ee
The effective two-dimensional action is given by
\be
S = \int d^2 x \sqrt{-g} e^{-2\Phi}  \left( R^{(2)} +2\p_i\Phi \p^i\Phi -2\p_i\varphi \p^i\varphi  -2\p_i\phi^I \p^i\phi^I \right)\ .
\label{dimred}
\ee
The solutions of this two-dimensional field theory
thus describe the gravitational interaction between four-dimensional plane waves which at the same time interact  with massless matter.
In particular, the plane waves may collide and gravitational collapse, in certain cases 
forming Killing-Cauchy horizons. 

The vacuum solutions are well known.
Our new results will include the construction of the most general solution to the system
in the presence of a non-zero stress tensor (including the back-reaction)
and  the interpretation of the gravitational dynamics
as scattering processes in two-dimensional gravity. In addition, we will identify a new class of four-dimensional geometries supported by massless scalars
free of curvature
singularities, which have apparently escaped the notice of earlier investigations.

In  contrast to other  exactly solvable gravitational models in two dimensions,
the present model originates
from  Einstein four-dimensional gravity;  for every phenomenon taking place in 
the two-dimensional world there is a
corresponding physical phenomenon in four dimensions.
Note  that other dimensional reduction ans\"atze from Einstein theory do not lead to a solvable model.
For example, the Einstein-Hilbert action minimally coupled to scalar fields and restricted to 
spherically symmetric configurations leads to equations that cannot be solved in general form.

It is convenient to choose the conformal gauge where the metric takes the familiar form,
\be
\label{gravio}
ds^2=- e^{2\sigma (x^+,x^-)} dx^+ dx^- + e^{-2\Phi(x^+,x^-)} (e^{2\varphi } dy^2 +e^{-2\varphi }  dz^2)\ ,\qquad \phi^I =\phi^I(x^+,x^-)\ ,
\ee
with $x^{\pm }=t\pm x$.  Except for the scalar fields, these are the gravitational waves  studied by Szekeres \cite{Szekeres:1972uu} and a particular case of Einstein-Rosen waves admitting Killing vectors
$\p_y$ and $\p_z$. One has a cylindrical wave upon interpreting $y$ as  azimuth angle and $x$ as  radial coordinate.
Using 
$R^{(2)}=8 e^{-2\sigma }\p_+\p_- \sigma $, 
 we find
\be
S = 2 \int dx^+ dx^- e^{-2\Phi}  \left( 2\p_+ \p_-\sigma   -2\p_+\Phi \p_-\Phi +2\p_+\phi^J \p_-\phi^J \right)\ .
\label{dimred2}
\ee
Here $J=1,...,N+1$ , with $\phi^{N+1}\equiv\varphi\ .$
By the following field redefinition, 
\be
u=e^{-2\Phi }\ ,\qquad v=-\sigma +\frac{\Phi}{2}\ ,
\ee
the action takes the form
\be
S = 4\int dx^+ dx^- \ \left( \p_+ u\p_- v    + u \p_+\phi^J \p_-\phi^J \right) \ .
\label{dimred3}
\ee
The action  (\ref{dimred3})  can alternatively be viewed as a 2d non-linear $\sigma $-model, 
where the $N+3$-dimensional target metric represents an homogeneous pp-wave  in 
Baldwin-Jeffery-Rosen coordinates 
(a detailed discussion on this two-dimensional  $\sigma $-model and its generalizations  can be found in \cite{Papadopoulos:2002bg}). A string $\sigma $-model
with action (\ref{dimred3}) and its connection with two-dimensional gravity first appeared in \cite{Tseytlin:1992va}. Here we are interested in the connection between (\ref{dimred3}) and four-dimensional Einstein gravity, which was not noticed before, and in aspects of the four-dimensional exact analytic solutions.

In terms of $u,\ v$, the  four-dimensional spacetime metric 
is 
\be
ds^2 =- \frac{e^{-2v}}{\sqrt{u} }\ dx^+dx^- + u (e^{2\varphi} dy^2+ e^{-2\varphi} dz^2)\ .
\nonumber
\ee


\noindent The key point of the solvability of the model is that the equation of motion for $v$ implies
\be
\p_+\p_- u=0\ \ \longrightarrow \ \ u=u_+(x^+)+u_-(x^-)\ .
\nonumber
\ee
 The remaining equations and constraints (coming from the $g_{++}$ and $g_{--}$ equations) are
 \bea
 \label{eqV}
 &&\p_+\p_- v= \p_+\phi^J \p_-\phi^J \ ,
 \\
\nonumber\\
\label{eqfi}
 &&\p_+(u\p_- \phi^J)+\p_-(u\p_+ \phi^J)=0\ ,
 \nonumber
 \\
\nonumber\\
 &&\p_\pm^2 u+ 2\p_\pm u\p_\pm v + 2u \p_\pm \phi^J \p_\pm \phi^J  =0\ .
 \nonumber
 \eea
By a conformal transformation we can choose $u$ to be linear in the coordinates.
This basically leads to three possible choices (with different physics).
%
%
%
The choice $ u=x^+$ leads to pp-waves with fields depending only on $x^+$.
The choices $u=2x$ or $u=2t$  involve more interesting physics.
In particular, $u=2t$ includes solutions representing the formation of
Milne horizons by colliding gravitational waves.  The case $u=2x$ includes domain wall singularities,
cylindrical gravitational waves and other types of travelling waves.
The  solutions in the two cases $u=2x$ or $u=2t$ are formally connected by suitably renaming coordinates. 
For concreteness, let us begin with the case
$u=2x $.
The equations of motion take the form  
\bea
&&  x (\p_t^2-\p_x^2)\phi^J - \p_x \phi^J=0\ ,
\label{besel}
\\
\nonumber\\
\label{eqV}
 &&(\p_t^2-\p_x^2) v= \p_t\phi^J \p_t\phi^J-\p_x\phi^J\p_x \phi^J \ ,
\eea
and the constraints become
\bea
&& \p_t v + 2x\p_t \phi^J \p_x \phi^J =0\ ,
\label{mas}
\\
\nonumber\\
&& \label{artan}
\p_x v +  x \left(\p_t\phi^J \p_t\phi^J+\p_x\phi^J\p_x \phi^J \right) =0\ .
\eea
Note that we have three equations for $v$. It is easy to check  that (\ref{mas}) and (\ref{artan}) are integrable and that they imply equation (\ref{eqV}).

The general solution to the classical  equations of motion including matter
is readily found by introducing the Fourier transform (we omit the superindex $J$)
\be
\phi(x,t) =b_0\ln x+ \int_{-\infty}^\infty d\omega\ e^{i \omega t} \tilde\phi (\omega,x)\ ,
\label{foura}
\ee
Thus $\td \phi $ obeys 
\be
 x (\omega^2+\partial_x^2)\td \phi + \p_x \td \phi=0\ ,
\nonumber
\ee
with general solution  
\be
\label{genia}
\td \phi(\omega,t) = a(\omega )\ J_0(\omega x) + b(\omega)\ Y_0(\omega x)\ ,
\ee 
where $J_0$ and $Y_0$ are  Bessel functions. 
Then, $v(x,t)$ is easily obtained by integration using (\ref{artan}).
This is a simple integral that leads to a quadratic form in Bessel functions, as shown below in an example.
This  describes all  solutions
 in GR of the form (\ref{gravio}) (adding to the family the similar solutions with $u=2t$ and the pp-waves with $u=x^+$). 

 The  solutions with $\phi^I=0$ are well known. 
 A classic example is   \cite{Szekeres:1972uu} 
 \be
 \varphi = \frac{1}{\sqrt{x^2-t^2}}\ ,\qquad v=\frac{x^2}{2(x^2-t^2)^2}\ ,\qquad \phi^I=0\ , 
 I=1,...,N\ .
\label{szek}
 \nonumber
 \ee
 It is easy to check that it solves (\ref{besel})--(\ref{artan}).
It can also be obtained by substituting $a(\omega)={\rm const.}$, $b(\omega )=0$, $b_0=0$,
  into the general solution (\ref{foura}), (\ref{genia}). 

\smallskip

One can study axisymmetric gravitational waves by renaming $x\to r$, $y\to \theta $, with $(r,\theta,z )$ representing
standard cylindrical coordinates.
Four-dimensional Minkowski spacetime ,
$ds^2=-dt^2+dr^2+r^2d\theta^2 +dz^2$,
arises from the vacuum solution $u=2r$,  $\varphi = \frac12 \ln \frac{r}{2}\ ,  v=-\frac14 \ln  2r$. 
 On the other hand, with the choice $u=2t$, the analogous solution describes a Milne space,
 $ds^2=-dt^2+dx^2+t^2dy^2 +dz^2$. These geometries will describe the geometry of
colliding plane waves near the focusing hypersurface for a specific choice of initial data
leading to a spacetime free from curvature singularities.

 A particular case of the general solution (\ref{genia}) corresponds to
 an axisymmetric wave  discussed by Kramer \cite{Kramer:1999tf}. 
 This is the vacuum solution  $\varphi =\frac12 \ln\frac{r}{2}+ a_0 J_0(\omega_0 x) \cos\w_0 t $. 
 This is a time-dependent solution which is regular at $r=0$. At this point,
  the Riemann-squared curvature invariant is a periodic function of $t$.
A more general class of regular vacuum solutions is found by setting
\be
 \varphi =\frac12 \ln\frac{r}{2}+ \int d\w\, a(\w )J_0(\omega r)  \cos\w t .
\ee
They represent standing cylindrical waves that extend the Kramer solution to more general wave profiles.
Another interesting solution is a standing  (but time-dependent) gravitational wave found in \cite{Gogberashvili:2009wa}
for Einstein theory coupled to a single massless scalar. The metric  has a domain wall singularity.

Strikingly, in the presence of a non-zero stress tensor, one can  find regular, {\it static}  geometries supported by massless scalar fields.
An example is the  solution
\be
\label{polin}
u=2r\ , \quad v=-\frac14 \ln  2r+\bar v(r)\ ,\quad \varphi = \frac12 \ln \frac{r}{2}\ , \quad \phi^1+i\phi^2 = A J_0(\w r) e^{i\w t}\ ,
\ee
with
\be
\bar v(r)= -A^2 r \omega  \left(r \omega ( J_0(r \omega ){}^2+ J_1(r \omega ){}^2)-J_1(r \omega ) J_0(r \omega
 ) \right) ,
\nonumber
\ee
leading to the geometry

\be
\label{polon}
ds^2 =e^{-2\bar v(r)}\ \big(-dt^2+dr^2\big)+ r^2 d\theta^2+ dz^2\ .
\ee

\noindent One can check that Einstein equations  (\ref{besel})--(\ref{artan}) are satisfied exactly.
This regular solution is new. The corresponding scalar curvature is 
\be
R^{(4)}=2 a^2 \omega ^2 \left(J_1(r \omega ){}^2-J_0(r \omega ){}^2\right) \exp \left[-2 a^2
   r \omega  \left(r \omega  \left(J_0(r \omega ){}^2+J_1(r \omega
   ){}^2\right)-J_0(r \omega ) J_1(r \omega )\right)\right].
\nonumber
\ee
Near $r=0$, the 2d metric approaches anti-de Sitter space with curvature $R^{(2)}=R^{(4)}\approx  -2a^2\omega^2+ O(r^2) $. As the radius is  increased, the curvature oscillates between negative and positive values with an amplitude that decays exponentially.
The solution admits a straightforward generalization by turning on other scalar fields with different frequencies, {\it viz.} $\phi^3+i\phi^4=A J_0(\w' r) e^{i\w' t}$, etc.

The analogous solution  corresponding to the choice $u=2t$ is formally obtained by  exchanging $r\leftrightarrow t$ in (\ref{polin}), (\ref{polon}) and it represents
an anisotropic  cosmology which is regular at $t=0$ where there is a Milne horizon. 
The scalar curvature flips sign with respect to the previous case and near $t=0$
the curvature approaches $+2a^2 \w^2$ (the two-dimensional metric approaches de Sitter space).
 The geometry can be analytically extended across the Milne horizon.

\smallskip

Colliding gravitational waves forming a Killing-Cauchy horizon 
can be described by the solution $u=2t$ and 
\be
 \varphi =\frac12 \ln\frac{t}{2}+ \int d\w\, J_0(\omega t)  \left(a_1(\w ) \cos\w x + a_2(\w )\sin\w x\right)\ .
\label{yurti}
\ee
 The horizons appear to be unstable under generic perturbations of the initial data
(a discussion can be found in \cite{Yurtsever:1988vc}).
The instability is not surprising, given that a generic perturbation of the form (\ref{genia}) is singular at $t=0$
 (note that the term  $b(\w )Y_0(\w t)$ contributes to
the singularity, due to the logarithmic behavior of the $Y_0(\w t)$ Bessel function near $t=0$). %

In the vacuum solution (\ref{yurti}) ``left" and ``right" moving modes 
have  the same amplitude, because  $\varphi$ is real.
Coupling GR to massless scalar fields  offers the possibility of studying  plane
wave collisions with  independent left and right-moving components,
  in terms of exact analytic
formulas that  include the back-reaction.
To this purpose, we define a complex field,
$\psi= \phi^1+i\phi^2$. For a mode of given frequency, one has
\be
 \psi^{\rm (L)}= A_\w\,  e^{i\w x} H_0^{(1)}(\omega t)\ , \qquad \psi ^{\rm (R)}= B_\w \, e^{i\w x}  H_0^{(2)}(\omega t )\ ,
\nonumber
\ee
where $ H_0^{(1,2)}$  are  Hankel functions.
At $t\to \pm\infty $, we have $\psi^{\rm (L,R)} \sim \frac{1}{\sqrt{ t}}  e^{i \omega(x \pm t)}$.
In two dimensions, these represent left-moving and right-moving particles propagating on the one dimensional space $x$.
The four-dimensional geometry is found by determining $v(x,t)$ through an elementary integration using (\ref{artan}), leading to a quadratic form in  the Hankel functions. Because of the logarithmic singularity in the Hankel functions, there is a space-like singularity at $t=0$, where the geometry approaches a Kasner solution.

An important issue concerns the extent to which the formation of singularities
in colliding gravitational plane waves may depend on the ultraviolet (UV) completion of the theory.
Classically, the focusing effect of plane waves makes colliding waves interact strongly and  produce singularities. The question is whether this can change if
the short-distance physics is governed by a different dynamics.
This is very difficult to address in four dimensions.
However, the present two-dimensional setup enables one to add
UV corrections to the action and at the same time preserve the solvability of the model.
An interesting model is constructed as follows:
one can add to action (\ref{dimred}) the local terms
$$
\Delta S= 2\alpha \int d^2 x \sqrt{-g} \left(\Phi R^{(2)} -
\p_i \Phi \p^i \Phi-2 \Phi \p_i \phi^J \p^i \phi^J\right)\ ,
$$
with  $\alpha>0 $. Because these terms are not multiplied by $e^{-2\Phi}$,
they have a negligible effect far away from the singularity where $e^{-2\Phi}\to\infty $.
The resulting two-dimensional gravity exhibits a number of interesting features and it is worth
to be investigated for its own sake.
In the conformal gauge, the resulting action still reduces to
the solvable model (\ref{dimred3}), this time with
\be
\label{polan}
u=e^{-2\Phi }+2\alpha \Phi\ ,\qquad v=-\sigma +\frac{\Phi}{2}\ .
\ee
$u$ is now bounded from below. 
The resulting physics is therefore not equivalent to the physics of the theory  (\ref{dimred})
(note that the field redefinition becomes singular at $du/d\Phi=0$).
The two-dimensional metric is now
$ds^2=- e^{-2v+\Phi(u)} dx^+dx^- $, where $\Phi(u)$ is the inverse field transformation
and one has to choose the branch that at large distances approaches $\Phi(u) \approx -\frac12 \ln u$,
where  plane waves are described by classical solutions of GR. 
However, at short distances, the new terms $\Delta S$ become important
and the singularity structure changes dramatically.
In the classical GR solutions, there is a singularity at $u=0$, where $\Phi\to +\infty$.
Now, as follows from (\ref{polan}), there is a
minimum value of $u$, $u_{\rm min}=\alpha \ln \frac{e}{\alpha}$, which is greater than zero 
for $\alpha< e $ and  the singularity at $u=0$ is never reached. 
In this case, $\Phi$ is bounded from above, $\Phi <\Phi_{\rm max} =-\frac12 \ln\alpha $
and $\Phi_{\rm max}$ represents a boundary of the spacetime.
By this  simple mechanism, the new terms $\Delta S$ prevent the formation
of the singularity. Physically,  the origin of  this 
phenomenon seems to be related to the fact that $\Phi $ freezes near $u_{\rm min}$, where its kinetic term gets multiplied by zero.
It is easy to write down the general exact solution to the equations of motion.
The solutions for $u$, $v$ and $\phi^J$
are formally given by the same general solution  described above, but the two-dimensional metric is different and
now the spacetime has a boundary at $\Phi=\Phi_{\rm max}$.
Constructing the full time evolution of the geometry therefore
requires a new physical input in terms of  boundary conditions.
It would be extremely interesting to explore
the  boundary dynamics in this theory, the evolution of the geometry and the global structure of the spacetime.

\smallskip
\bigskip


We are grateful to A. Tseytlin for useful comments.
We acknowledge financial support from projects  FPA2013-46570,  
 2017-SGR-929  and  MDM-2014-0369 of ICCUB (Unidad de Excelencia `Mar\'\i a de Maeztu').



\begin{thebibliography}{20}

\bibitem{Einstein:1937qu} 
  A.~Einstein and N.~Rosen,
  ``On Gravitational waves,''
  J.\ Franklin Inst.\  {\bf 223}, 43 (1937).

\bibitem{Bondi:1958aj}
  H.~Bondi, F.~A.~E.~Pirani and I.~Robinson,
  ``Gravitational waves in general relativity. 3. Exact plane waves,''
  Proc.\ Roy.\ Soc.\ Lond.\ A {\bf 251} (1959) 519.

\bibitem{Khan:1971vh}
  K.~A.~Khan and R.~Penrose,
  ``Scattering of two impulsive gravitational plane waves,''
  Nature {\bf 229} (1971) 185.


\bibitem{Szekeres:1972uu}
  P.~Szekeres,
  ``Colliding plane gravitational waves,''
  J.\ Math.\ Phys.\  {\bf 13} (1972) 286.


\bibitem{Chandrasekhar:1986jn}
  S.~Chandrasekhar and B.~C.~Xanthopoulos,
  ``A New Type of Singularity Created by Colliding Gravitational Waves,''
  Proc.\ Roy.\ Soc.\ Lond.\ A {\bf 408} (1986) 175.


\bibitem{Yurtsever:1988vc}
  U.~Yurtsever,
  ``Structure of the Singularities Produced by Colliding Plane Waves,''
  Phys.\ Rev.\ D {\bf 38} (1988) 1706.

\bibitem{Feinstein:1989hg} 
  A.~Feinstein and J.~Ibanez,
  ``Curvature Singularity - Free Solutions for Colliding Plane Gravitational Waves With Broken U - $V$ Symmetry,''
  Phys.\ Rev.\ D {\bf 39}, 470 (1989).

\bibitem{Griffiths:1991zp}
  J.~B.~Griffiths,
  ``Colliding plane waves in general relativity,''
  Oxford, UK: Clarendon (1991) 232 p. (Oxford mathematical monographs).

\bibitem{Kramer:1999tf} 
 D.~Kramer,
  ``Exact gravitational wave solution without diffraction,''
  Class.\ Quant.\ Grav.\  {\bf 16}, L75 (1999).

\bibitem{Stephani:2003tm}
  H.~Stephani, D.~Kramer, M.~A.~H.~MacCallum, C.~Hoenselaers and E.~Herlt,
  ``Exact solutions of Einstein's field equations' (Cambridge University press, Cambridge 2003).


 





\bibitem{Papadopoulos:2002bg}
   G.~Papadopoulos, J.~G.~Russo and A.~A.~Tseytlin,
  ``Solvable model of strings in a time dependent plane wave background,''
 Class.\ Quant.\ Grav.\  {\bf 20} (2003) 969.


\bibitem{Tseytlin:1992va}
  A.~A.~Tseytlin,
  ``A Class of finite two-dimensional sigma models and string vacua,''
  Phys.\ Lett.\ B {\bf 288} (1992) 279.


  \bibitem{Gogberashvili:2009wa}
   M.~Gogberashvili, S.~Myrzakul and D.~Singleton,
   ``Standing gravitational waves from domain walls,''
   Phys.\ Rev.\ D {\bf 80} (2009) 024040.



\end{thebibliography}
\end{document}